\documentclass[12pt]{article}
\usepackage{amsfonts}
\usepackage{latexsym}
\title{Asymptotics of eigenvalues of the operator describing Aharonov-Bohm effect combined  with homogeneous magneticfield coupled with a strong  $\delta$-interaction on a loop}
\newlength{\dinwidth}
\newlength{\dinmargin}
\def\ben{\begin{displaymath}}
\def\een{\end{displaymath}}
\def\beq{\begin{equation}}
\def\eeq{\end{equation}}
\def\beqs{\begin{displaymath}}
\def\eeqs{\end{displaymath}}
\def\beqn{\begin{eqnarray}}
\def\eeqn{\end{eqnarray}}

\def\R{\mathbb {R}}
\def\N{\mathbb{N}}

\newtheorem{thm}{Theorem}[section]

\newtheorem{prop}{Proposition}[section]
\newtheorem{coro}{Corollary}[section]
\newtheorem{lem}{Lemma}[section]

 \def \R {\Bbb R}
 \def \N {\Bbb N }

\begin{document}
\title{Asymptotics of eigenvalues of the operator describing Aharonov-Bohm effect combined  with homogeneous magnetic field coupled with a strong  $\delta$-interaction on a loop}

\author{G.~Honnouvo$^{a,b}$ and M. N.~Hounkonnou$^{b}$}
\date{}
\maketitle
\maketitle
\begin{quote}
{\small \em a) Department of Mathematics and Statistics,\\
\phantom{e)x}Concordia University,\\
\phantom{e)x}7141 Sherbrooke Street West Montreal, Quebec
Canada H4B 1R6\\
b)Unit\'e de Recherche en Physique Th\'eorique (URPT)\\
\phantom{e)x}Institut de Math\'ematiques
et de Sciences  Physiques (IMSP) \\
\phantom{e)x}01 B.P. 2628 Porto-Novo, Benin, \\
 \phantom{e)x}International Chair in Mathematical Physics and Applications (ICMPA)\\
\phantom{e)x}01 BP 2628 Porto-Novo, Benin\\
 \rm \phantom{e)x}g${}_{-}$honnouvo@yahoo.fr, hounkonnou1@yahoo.fr}

\vspace{8mm}

\noindent {\small \begin{abstract}
We investigate the two-dimensional magnetic operator
 $H_{c_0,B,\beta} = {(-i\nabla -A)}^{2}-\beta\delta(.-\Gamma),$ where 
$\Gamma$ is a smooth loop. The vector potential has the form $A=c_0\bigg(\frac{-y}{{x^2+y^2}}; 
\frac{x}{{x^2+y^2}}    \bigg)+ \frac{B}{2}\bigg(-y; 
x\bigg) $; $B>0,$ $c_0\in]0;1[$. The asymptotics of negative eigenvalues of 
$H_{c_0,B,\beta}$ for $\beta \longrightarrow +\infty$ is found. We also prove that 
for large enough positive value of $\beta$ the system exhibits persistent currents.
\end{abstract}}
\end{quote}

\section{Introduction}
In the presence of a static magnetic field, a single isolated 
normal-metal loop is predicted to carry an equilibrium current\cite{BIL}, which 
is periodic in the magnetic flux $\Phi$ threading the loop. This current 
arises due to the boundary conditions \cite{BY}
imposed by the doubly connected nature of the loop. As a consequence of 
these boundary conditions, the free energy $E$ and the thermodynamic 
current $I(\Phi) = \frac{\partial E}{\partial \Phi}$ are periodic in 
$\Phi,$ with a fundamental period $\Phi_0 = \hbar /e .$ 
In recent papers \cite{EY1} '\cite{EY2} Exner and Yoshitomi have 
derived an asymptotic formula showing that if the $\delta-$coupling is strong 
or in a homogeneous magnetic fied $B$ perpendicular to the plane, the 
negative eigenvalues approach those of the ideal model in which the 
geometry of $\Gamma$ is taken into account by means of an effective 
curvature-induced potential. In \cite{hh}, the same result is proved when the vector potential is the Aharonov one.

The pupose of this paper is to prove the same result 
when the electron is subject to a Bohm-Aharonov effect in the plane on the background of a homogeneous magnetic field. As a consequence of this result, we prove that the system exhibits persistent currents.

\section{The model and the results}
In this section, we study  the magnetic operator in 
$L^2(\R^2)$ with an attractive $\delta$-interaction applied to a loop. We use 
the gauge fied $A = c_0\bigg(\frac{-y}{{x^2+y^2}}; 
\frac{x}{{x^2+y^2}}    \bigg)+ \frac{B}{2}\bigg(-y; 
x\bigg) .$

Let $ \Gamma :[0,L] \ni s \mapsto (\Gamma_1(s), \Gamma_2(s)) \in 
\R^2$ be the closed counter-clockwise $C^4$ Jordan curve which is parametrized 
by its arc length. Given $\beta >0$ and $(c_0,B) \in ]0,1[\times]0, \infty[,$ we define the 
quadratic form

\begin{eqnarray}
q_{c_0,B,\beta}(f;f)&=&||(-i\partial _x + \frac{c_0y}{{x^2+y^2}} 
+\frac{B}{2}y)f||^2_{L^2(\R^2)}\nonumber
\\
& +& ||(-i\partial _y - \frac{c_0x}{{x^2+y^2}}-\frac{B}{2}x)f||^2_{L^2(\R^2)}    
-\beta\int_\Gamma |f(x)|^2ds 
\end{eqnarray}

with the domain $H^1(\R^2)$, where $\partial_x \equiv 
\frac{\partial}{\partial_x}$, and the norm refers to $L^2(\R^2).$

Let us denote by $H_{c_0,B,\beta}$ the self-adjoint operator associated 
to the form $q_{c_0,B,\beta}(\:,\:)$:

$$ H_{c_0,B,\beta} = {(-i\nabla -A)}^{2}-\beta\delta(.-\Gamma).$$
Our main goal is to study, as in \cite{EY2}, the asymptotic behaviour of the negative 
eigenvalues of $H_{c_0,B,\beta}$
as $\beta \longrightarrow +\infty.$ 

Let $\gamma :\R   \longrightarrow \R$
be the signed curvature of $\Gamma$ , i.e.
$$ \gamma(s):= \bigg( 
\Gamma_1''\Gamma_2'-\Gamma_2''\Gamma_1'\bigg)(s).$$

Next we need a comparison operator on the cuve
\begin{eqnarray}
S_{c_0,B} = -\frac{d^2}{ds^2}- \frac{1}{4}\gamma(s)^2\: in\: L^2\bigg((0; L)\bigg),
\end{eqnarray}

with the domain 
\begin{eqnarray}
P_{c_0,B} = \{ u\in H^2(]0; L[); u^{(k)}(L) = u^{(k)}(0);\:k= 1,2 \}.
\end{eqnarray}
For $j\in\N,$ we denote by $\mu_j(c_0,B)$ the $j^{th}$ eigenvalue of the 
operator $S_{c_0,B}$ counted with multiplicity. This allows us to formulate 
our main result, and his proof follows in the same way as in \cite{EY2}:

\smallskip\begin{thm}\label{t1} 
Let $n$ be an arbitrary integer and $I$ be a nonempty compact subset of $]0,1[\times]0, \infty[$. Then  there exists 
$\beta (n, I)$ such that $\# \bigg\{ \sigma_{d}(H_{c_0,B,\beta })\cap 
]-\infty, 0[ \bigg\}\geq n$  for $\beta\geq \beta(n,I)$ and $(c_0,B) \in I.$ 

For $\beta\geq \beta(n, I)$ and $(c_0,B) \in I$ we denote by 
$\lambda_n({c_0,B,\beta})$ the $n^{th}$ eigenvalue of $H_{c_0,B, \beta}$ counted with 
multiplicity.

Then $\lambda_n({c_0,B, \beta })$ admits an asymptotic expansion of the 
form

$\lambda_n({c_0,B, \beta }) = -\frac{1}{4}\beta^2 + \mu_n(c_0,B) + \mathcal{O} 
(\beta^{-1}\ln \beta)$ as $\beta\rightarrow +\infty$; where the error term 
is uniform with respect to $(c_0,B)\in I.$
\end{thm}\smallskip

The existence of persistent currents is given by the  
consequence of the following result.

\smallskip\begin{coro}\label{c1} 
Let $n\in \N.$ Then there exists a constant $\beta_1(n, 
I)>0$ such that the function $\lambda_n({.,.,\beta})$ is not constant for 
$\beta >\beta_1(n, I) .$
\end{coro}\smallskip

Since the spectral properties of $H_{c_0,B,\beta}$ are cleary invariant 
with respect to Euclidean transformation of the plane, we may assume 
without any loss of generality that the curve $\Gamma$ parametrized in the 
following way:

$\Gamma_1(s) = \Gamma_1(0) + \int_0^s \cos H(t) dt \qquad \Gamma_2(s) = 
\Gamma_2(0) + \int_0^s \sin H(t) dt$

where $H(t) \equiv -\int_0^t \gamma (u)du .$ Le $\Psi_a$ be the map

$ \Psi_a :[0,L)\times(-a,a)\ni(s,u)\mapsto (\Gamma_1(s)-u\Gamma_2'(s), 
\Gamma_2(s)+u\Gamma_1'(s) ) \in \R^2.$

From \cite{EY1} we know that there exists $a_1>0$ such that the map 
$\Psi_a$ is injective for all $a\in (0,a_1].$ We thus fix $a\in(0,a_1)$ 
and denote by $\Sigma_a$ the strip of width $2a$ enclosing $\Gamma.$

$$\Sigma_a \equiv \Psi_a([0,L)\times (-a,a)).$$

Then the set $\R^2/{\Sigma_a}$ consists of two connected components 
which we denote by $\wedge_a ^{in}$ and $\wedge_a ^{out},$ where the 
interior one, $\wedge_a ^{in},$ is compact. We define a pair of quadratic 
forms,

\begin{eqnarray}
q_{c_0 ,B,a, \beta} ^{\pm}(f;f)&=&||(-i\partial _x + \frac{c_0y}{{x^2+y^2}}+ 
By)f||^2_{L^2(\Sigma_a)}\\
& +& ||(-i\partial _y - 
\frac{c_0x}{{x^2+y^2}}-Bx)f||^2_{L^2(\Sigma_a)}    -\beta\int_\Gamma |f(x)|^2ds \nonumber
\end{eqnarray}

which are given by the same expression but differ by their domains, the 
latter in $H_0^1(\Sigma_a)$ for $q_{c_0 ,a, \beta} ^{+}$ and $H^1(\Sigma_a)$ for 
$q_{c_0 ,a, \beta} ^{-}.$ Furthermore, we introduce the quadratic forms

\begin{eqnarray}
e_{c_0,B a} ^{\pm}(f;f)&=&||(-i\partial _x + \frac{c_0y}{{x^2+y^2}} 
+By)f||^2_{L^2(\wedge_a ^j)}\\
& +& ||(-i\partial _y - 
\frac{c_0x}{{x^2+y^2}}-Bx)f||^2_{L^2(\wedge_a ^j)} 
\end{eqnarray}

for $j= out,\: in,$ with the domain $H_0^1(\wedge_a ^j)$ and  
$H^1(\wedge_a ^j)$ corresponding to the $\pm$ sign respectively. Let $L_{c_0 ,B, 
a, \beta}^{\pm},\: E_{c_0,B, a}^{out, \pm}$ and $ E_{c_0,B, a}^{in, \pm}$ 
be the self-adjoint operators associated with the forms $q_{c_0,B,a, \beta 
}^{\pm}, \: e_{c_0,B, a} ^{out,\pm}$ and $e_{c_0,B, a} ^{in,\pm},$ 
respectively.

As in \cite{EY1} we are going to use the dirichlet-Neumann bracketing 
with additional boundary conditions at the boundary of $\Sigma_a.$ One can easily see this by comparing the form domains of the 
involved operators, cf[\cite{EY2} or (\cite{rs}, thm XIII.2)]. We get

\begin{eqnarray}\label{ok}
E_{c_0,B, a}^{in,-}\oplus L_{c_0 ,B, a, \beta}^{-} \oplus E_{c_0,B, 
a}^{out,-} \leq H_{c_0,B,a}\leq E_{c_0,B, a}^{in,+}\oplus L_{c_0 ,B, a, \beta}^{+} 
\oplus E_{c_0, a}^{out,+}    
\end{eqnarray}
with the decomposed estimating operators in $L^2(\R^2) = L^2(\wedge_a 
^{in})\oplus L^2(\Sigma_a )\oplus L^2(\wedge_a ^{out}).$ In order to 
assess the negative eigenvalues of $H_{c_0,B, \beta},$ it suffices to 
consider those of $L_{c_0 ,B, a, \beta}^{+}$ and $L_{c_0 ,B, a, \beta}^{-},$ 
because the other operators involved in $(\ref{ok})$ are positive. Since the loop 
is smooth, we can pass inside $\Sigma_a$ to the natural curvilinear 
coordinates. We state

$$(U_af)(s,u) = {(1 + u\gamma(s))}^{1/2}f(\Psi_a(s,u))\qquad 
\mbox{for}\: f\in L^2(\Sigma_a )$$

which defines the unitary operator $U_a$ from $L^2(\Sigma_a )$ to 
$L^2((0,L )\times(-a,a)).$ To express the estimating operators in the new 
variables, we introduce

${\cal{Q}} _a^+  = \bigg\{ \psi\in H^1((0,L )\times(-a,a));\:\psi(L,.) 
=\psi(0,.)on (-a,a);\quad \psi(.,a) =\psi(.,-a) on  (0,L)  \bigg\}$

${\cal{Q}} _a^- =  \bigg\{ \psi\in H^1((0,L )\times(-a,a));\quad \psi(L,.) 
=\psi(0,.)\quad on \quad(-a,a) \bigg\}  $

and define the quadratic forms

\begin{eqnarray}\label{ba4}
z_{c_0,B,a,\beta}^\pm [g] &=&\int_0^L \int_{-a}^a {(1 + 
u\gamma(s))}^{-2}|\partial_s g|^2 duds + \int_0^L \int_{-a}^a|\partial_u g|^2 \nonumber 
\\
&+& \int_0^L \int_{-a}^a V(s,u) |g|^2 dsdu - \beta \int_0^L|g(s,0)|^2ds 
\nonumber \\
& -& \frac{b_\pm}{2}\int_0^L \frac{\gamma (s)}{1 + a \gamma 
(s)}|g(s,a)|^2ds + \frac{b_\pm}{2}\int_0^L \frac{\gamma (s)}{1 - a \gamma 
(s)}|g(s,-a)|^2ds \nonumber \\
&+& \int_0^L \int_{-a}^a\theta (s,u)^{-1}\left(c_0\theta (s,u)+\frac{1}{2}B \right)^2 |g|^2duds \nonumber \\
&+& Im \int_0^L \int_{-a}^a \bigg(   2c_0 \theta (s,u)  + B   \bigg)(\Gamma_2  + u 
\Gamma_1')\bigg({(1+ u\gamma)}^{-1}\cos H \overline{g}\partial_s g - \sin H 
\overline{g}\partial_u g  \bigg) duds \nonumber \\
&-& Im \int_0^L \int_{-a}^a \bigg(   2c_0 \theta (s,u)  + B   \bigg)(\Gamma_1  - u 
\Gamma_2')\bigg({(1+ u\gamma)}^{-1}\sin H \overline{g}\partial_s g + \cos H 
\overline{g}\partial_u g  \bigg) duds\nonumber\\
\end{eqnarray}

on ${\cal{Q}}_a^\pm $ respectively, where

$V(s,u) = 1/2{(1 + u\gamma (s))}^{-3}u \gamma(s)'' -5/4{(1 + u\gamma 
(s))}^{-4}u^2\gamma'(s)^2 -1/4{(1 + u\gamma (s))}^{-2}\gamma (s)^2 , $ 

$\theta (s,u) = {\bigg( \Gamma_1^2(s) + \Gamma_2^2(s) + u^2 - 2u 
\big(\Gamma_1(s)\Gamma_2'(s) -\Gamma_2(s)\Gamma_1'(s) \big) \bigg) }^{-1}$

$b_+=0$ and $b_- = 1.$

(\ref{ba4}) can be write as

\begin{eqnarray}\label{ba12}
b_{c_0,B,a,\beta}^\pm [g] &=&\int_0^L \int_{-a}^a {(1 + 
u\gamma(s))}^{-2}|\partial_s g|^2 duds + \int_0^L \int_{-a}^a|\partial_u g|^2 \nonumber 
\\
&+& \int_0^L \int_{-a}^a V(s,u) |g|^2 dsdu - \beta \int_0^L|g(s,0)|^2ds 
\nonumber \\
& -& \frac{b_\pm}{2}\int_0^L \frac{\gamma (s)}{1 + a \gamma 
(s)}|g(s,a)|^2ds + \frac{b_\pm}{2}\int_0^L \frac{\gamma (s)}{1 - a \gamma 
(s)}|g(s,-a)|^2ds \nonumber \\
&+& \int_0^L \int_{-a}^a\alpha_1 (s,u) |g|^2duds \nonumber \\
&+& Im \int_0^L \int_{-a}^a \alpha_2 (s,u)\bigg({(1+ u\gamma)}^{-1}\cos H \overline{g}\partial_s g - \sin H 
\overline{g}\partial_u g  \bigg) duds \nonumber \\
&-& Im \int_0^L \int_{-a}^a \alpha_3 (s,u)\bigg({(1+ u\gamma)}^{-1}\sin H \overline{g}\partial_s g + \cos H 
\overline{g}\partial_u g  \bigg) duds\nonumber\\
&=& \int_0^L \int_{-a}^a {(1 + 
u\gamma(s))}^{-2}|\partial_s g|^2 duds + \int_0^L \int_{-a}^a|\partial_u g|^2 \nonumber 
\\
&+& \int_0^L \int_{-a}^a V(s,u) |g|^2 dsdu - \beta \int_0^L|g(s,0)|^2ds 
\nonumber \\
& -& \frac{b_\pm}{2}\int_0^L \frac{\gamma (s)}{1 + a \gamma 
(s)}|g(s,a)|^2ds + \frac{b_\pm}{2}\int_0^L \frac{\gamma (s)}{1 - a \gamma 
(s)}|g(s,-a)|^2ds \nonumber \\
&+& \int_0^L \int_{-a}^a\alpha_1 (s,u) |g|^2duds \nonumber \\
&+& Im \int_0^L \int_{-a}^a\bigg( \alpha_2 (s,u)\cos H-\alpha_3 (s,u)\sin H\bigg){(1+ u\gamma)}^{-1}(\overline{g}\partial_s g )duds \nonumber \\
&-& Im \int_0^L \int_{-a}^a \bigg( \alpha_3 (s,u)\cos H-\alpha_2 (s,u)\sin H\bigg){(1+ u\gamma)}^{-1}(\overline{g}\partial_u g) duds\nonumber
\end{eqnarray}

where

$\alpha_1 (s,u) = \theta (s,u)^{-1}\left(c_0\theta (s,u)+\frac{1}{2}B \right)^2 ,$

$\alpha_2 (s,u) =  \bigg(   2c_0 \theta (s,u)  + B   \bigg)(\Gamma_2  + u 
\Gamma_1' )$ and

$\alpha_3 (s,u) =  \bigg(   2c_0 \theta (s,u)  + B   \bigg)(\Gamma_1  - u 
\Gamma_2')$

Let $D_{c_0,B, a,\beta}^{\pm}$ be the self-adjoint operators associated 
with the forms 
$b_{c_0,B,a,\beta}^\pm $, respectively. By analogy with \cite{EY1}, we 
get the following result.

\smallskip\begin{lem}\label{1} 
$U_a D_{c_0,B, a,\beta}^{\pm}U_a = L_{c_0,B, a,\beta}^{\pm}.$
\end{lem}\smallskip

In order to eliminate the coefficients of $\overline{g}\partial_s g$ 
and $\overline{g}\partial_u g$ in (\ref{ba4}) modulo small erros, we 
employ the following unitary operator

\begin{eqnarray}\label{up}
(M_{c_0,B}h)(s,u) = \exp [iK(s,u)]h(s,u).
\end{eqnarray}

Replacing $M_{c_0,B}h$ in (\ref{ba4}), it becomes:

\begin{eqnarray}\label{ba1}
v_{c_0,B,a,\beta}^\pm [g]&=& \int_0^L \int_{-a}^a {(1 + 
u\gamma(s))}^{-2}|\partial_s g|^2 duds + \int_0^L \int_{-a}^a|\partial_u g|^2 \nonumber 
\\
&+& \int_0^L \int_{-a}^a V(s,u) |g|^2 dsdu - \beta \int_0^L|g(s,0)|^2ds 
\nonumber \\
& -& \frac{b_\pm}{2}\int_0^L \frac{\gamma (s)}{1 + a \gamma 
(s)}|g(s,a)|^2ds + \frac{b_\pm}{2}\int_0^L \frac{\gamma (s)}{1 - a \gamma 
(s)}|g(s,-a)|^2ds \nonumber \\
&+& \int_0^L \int_{-a}^a\bigg(\alpha_1 (s,u)+|K_s|^2 {(1 + 
u\gamma(s))}^{-2} +|K_u|^2 +K_s\Omega_1(s,u)-K_u\Omega_2(s,u)  \bigg)|g|^2duds \nonumber \\
&+& Im \int_0^L \int_{-a}^a\bigg(\Omega_1(s,u)+ 2K_s{(1 + 
u\gamma(s))}^{-2}   \bigg)(\overline{g}\partial_s g )duds \nonumber \\
&-& Im \int_0^L \int_{-a}^a\bigg(\Omega_2(s,u)-2K_u     \bigg)(\overline{g}\partial_u g) duds
\end{eqnarray}

where 
\begin{eqnarray}
\Omega_1(s,u) =\bigg( \alpha_2 (s,u)\cos H-\alpha_3 (s,u)\sin H\bigg){(1+ u\gamma)}^{-1} ,\\
\Omega_2(s,u)=\bigg( \alpha_3 (s,u)\cos H-\alpha_2 (s,u)\sin H\bigg){(1+ u\gamma)}^{-1}\\
K_s=\frac{\partial_s K(s,u)}{\partial s},\:\:K_u=\frac{\partial_u K(s,u)}{\partial u}\:\:g_s=\frac{\partial_s g(s,u)}{\partial s},\:\:g_u=\frac{\partial_u g(s,u)}{\partial u}
\end{eqnarray}

To eliminate the coefficients of $\overline{g}\partial_u g$ in $c_{c_0,B,a,\beta}^\pm [g] $, we have the following differential equation:

 \begin{eqnarray}
\frac{\partial K(s,u)}{\partial_u}=\frac{1}{2}\Omega_2(s,u) 
\end{eqnarray}

and then, we have

\begin{eqnarray}
K(s,u)=\frac{1}{2}\int_0^u \Omega_2(s,v) dv.
\end{eqnarray}

This form of $K,$ reduces (\ref{ba1}) to:

\begin{eqnarray}\label{b}
\tilde b_{c_0,B,a,\beta}^\pm [g]&=& \int_0^L \int_{-a}^a {(1 + 
u\gamma(s))}^{-2}|\partial_s g|^2 duds + \int_0^L \int_{-a}^a|\partial_u g|^2 \nonumber 
\\
&+& \int_0^L \int_{-a}^a V(s,u) |g|^2 dsdu - \beta \int_0^L|g(s,0)|^2ds 
\nonumber \\
& -& \frac{b_\pm}{2}\int_0^L \frac{\gamma (s)}{1 + a \gamma 
(s)}|g(s,a)|^2ds + \frac{b_\pm}{2}\int_0^L \frac{\gamma (s)}{1 - a \gamma 
(s)}|g(s,-a)|^2ds \nonumber \\
&+& \int_0^L \int_{-a}^a\bigg(\alpha_1 (s,u)+|K_s|^2 {(1 + 
u\gamma(s))}^{-2} +|K_u|^2 +K_s\Omega_1(s,u)-K_u\Omega_2(s,u)  \bigg)|g|^2duds \nonumber \\
&+& Im \int_0^L \int_{-a}^a\bigg(\Omega_1(s,u)+ 2K_s{(1 + 
u\gamma(s))}^{-2}   \bigg)(\overline{g}\partial_s g )duds 
\end{eqnarray}

for $g\in  {\cal{Q}} _a^\pm ,$ respectively.

Let $\tilde D_{c_0,B,a,\beta}$ be the self-adjoint operators associated with the 
forms $\tilde b_{c_0,B,a,\beta}^\pm ,$ respectively. We have the 
following result

\smallskip\begin{lem}\label{2} 
$M_{c_0,B}^* D_{c_0,B, a,\beta}^{\pm}M_{c_0,B} = \tilde D_{c_0,B, 
a,\beta}^{\pm}.$
\end{lem}\smallskip

 In the estimation of the $\tilde D_{c_0,B, a,\beta}^{\pm}$, let us use 
the same notation as  in \cite{EY2}:

$\gamma_+ =   \displaystyle{\max_{[0,L]}}| \gamma(.)|$

$N_{c_0,B}(a)=\displaystyle{ \max_{(s,u)\in [0,L]\times[-a,a]}} |\bigg(\Omega_1(s,u)+ 2K_s{(1 + 
u\gamma(s))}^{-2}   \bigg)|$
 
and

$M_{c_0,B}(a) :=\displaystyle{ \max_{(s,u)\in [0,L]\times[-a,a]}} 
|W_{c_0,B}(s,u)+\frac{1}{4}\gamma (s)^2 |;$ where

\begin{eqnarray}\label{ab}
W_{c_0,B}(s,u)&=&V(s,u)+\alpha_1 (s,u)+|K_s|^2 {(1 + 
u\gamma(s))}^{-2} +|K_u|^2\nonumber\\ 
&+& K_s\Omega_1(s,u)-K_u\Omega_2(s,u).
\end{eqnarray}

Since $(c_0,B) \in I$, then there exists 
$T$ such that 

$N_{c_0,B}(a) + M_{c_0,B}(a) \leq Ta \qquad $ for $0< 
a<\frac{1}{2\gamma_+}$ and  $ (c_0,B) \in I,$
where $T$ is independent of $a,\:c_0$ and $B$. For fixed $0< 
a<\frac{1}{2\gamma_+},$ as in \cite{EY2} we  define 

\begin{eqnarray}\label{b}
\hat b_{c_0,B,a,\beta}^\pm [g] &=&\int_0^L \int_{-a}^a \bigg(\bigg[ {(1 
\pm u\gamma_+)}^{-2} \pm \frac{1}{2} N_{c_0,B}(a)\bigg]|\partial_s g|^2  + 
|\partial_u g|^2 \nonumber \\
&+& \bigg[ - \frac{1}{4} \gamma(s)^2\pm \frac{1}{2}N_{c_0,B}(a) \pm 
M_{c_0,B}(a)   \bigg]|g|^2\bigg) duds \nonumber \\
&-& \beta \int_0^L|g(s,o)|^2ds - \gamma_+ b_\pm\int_0^L 
\bigg(|g(s,a)|^2 + |g(s,-a)|^2 \bigg)ds 
\end{eqnarray}

for $g\in {\cal{Q}} _a^\pm ,$ respectively. Since $|Im(\overline 
g\partial _sg)|\leq \frac{1}{2} \bigg(|g|^2 + |\partial_s g|^2\bigg),$ 
we obtain

\begin{eqnarray}\label{ac1}
\tilde b_{c_0,B,a,\beta}^+[g] \leq \hat b_{c_0,B,a,\beta}^+ [g] \qquad 
\mbox{for}\qquad \:  g\in {\cal{Q}} _a^+ 
\end{eqnarray}

\begin{eqnarray}\label{ac2}
\hat b_{c_0,B,a,\beta}^-[g] \leq \tilde b_{c_0,B,a,\beta}^- [g] \qquad 
\mbox{for}\qquad \:  g\in {\cal{Q}} _a^- .
\end{eqnarray}

Let $\hat H^\pm_{c_0,B,a,\beta}$ be the self-adjoint operators associated 
with the form 
$\hat b_{c_0,B,a,\beta}^\pm ,$ respectively.

Furthermore, let $T^+_{a,\beta}$ be the self-adjoint operator 
associated with the form

$t^+_{a,\beta}[f] = \int_{-a}^a |f'(u)|^2du - \beta|f(0)|^2; \qquad f\in H_0^1(]-a,a[),$ 

and similarly, let $T^-_{a,\beta}$ be the self-adjoint operator 
associated with the form

$t^-_{a,\beta}[f] = \int_{-a}^a |f'(u)|^2du - \beta|f(0)|^2- 
\gamma_+\big(|f(a)|^2+ |f(-a)|^2\big); \qquad f\in H^1(]-a,a[).$

As in \cite{EY2}, let us denote by $\mu_j^\pm(c_0,B,a)$ the $j^{th}$ 
eigenvalue of the following operator, define on $L^2(]0,L[)$, by 

\begin{eqnarray}
U^\pm_{a,\beta} = -\bigg[ {(1 \mp u\gamma_+)}^{-2} \pm \frac{1}{2} 
N_{c_0,B}(a)\bigg]\frac{d^2}{ds^2} - \frac{1}{4} \gamma(s)^2\pm 
\frac{1}{2}N_{c_0,B}(a) \pm M_{c_0,B}(a) 
\end{eqnarray}

in $L^{2}((0,L))$ with the domain $P_{c_0}$ specified in the
 previous section. Then we have
 \begin{equation} \label{decomp}
\hat{H}^{\pm}_{c_0,B,a,\beta}= U_{c_0,B,a}^{\pm}\otimes 1+1\otimes
T^{\pm}_{a,\beta}.
 \end{equation}
Let $\mu_{j}^{\pm}(c_0,B,a)$ be the $j$-th eigenvalue of
 $U^{\pm}_{c_0,B,a}$ counted with multiplicity. We shall prove the
 following estimate as in \cite{EY2}.
\begin{prop}\label{EST1}
Let $j\in \N.$ Then there exists $C(j)>0$ such that \\
$|\mu_j^+(c_0,B,a) -  \mu_j(c_0,B) | + | \mu_j^-(c_0,B,a) -  \mu_j(c_0,B) |\leq 
C(j)a$\\
holds for $(c_0,B)\in I$ and $0<a<\frac{1}{2\gamma_+},$ where $C(j)$ is 
independent of $c_0,\: B$ and $a.$
\end{prop}

{\bf{Proof:}}

{\sl Proof:} Since
 \begin{eqnarray*} 
\lefteqn{ U^{+}_{c_0,B,a} -\left\lbrack
(1-a\gamma_{+})^{-2}+\frac{1}{2}N_{c_0,B}(a) \right\rbrack S_{c_0,B}} \\
&& = \frac{1}{4}\left\lbrack
\frac{a\gamma_{+}(2-a\gamma_{+})}{(1-a\gamma_{+})^{2}}
+\frac{1}{2}N_{c_0,B}(a) \right\rbrack
\gamma(s)^{2}+\frac{1}{2}N_{c_0,B}(a)+M_{c_0,B}(a)\,,
 \end{eqnarray*}
 $N_{c_0,B}(a)+M_{c_0,B}(a)\leq Ta$ for $0<a<
\frac{1}{2\gamma_{+}}$ and $(c_0,B)\in I$, we infer that there is a
 constant $C_{1}>0$ such that
 $$ 
\left\Vert U^{+}_{c_0,B,a} - \left\lbrack
(1-a\gamma_{+})^{-2}+\frac{1}{2}N_{c_0,B}(a) \right\rbrack S_{c_0,B}
\right\Vert \leq C_{1}a
 $$ 
for $0<a<\frac{1}{2\gamma_{+}}$ and $(c_0,B)\in I$. This together with
 the min-max principle implies that
 $$ 
\left|\mu^{+}_{j}(c_0,B,a)- \left\lbrack
(1-a\gamma_{+})^{-2}+\frac{1}{2}N_{c_0,B}(a) \right\rbrack \mu_{j}(c_0,B)
\right | \leq C_{1}a
 $$ 
for $0<a<\frac{1}{2\gamma_{+}}$ and $(c_0,B)\in I$. Since
 $\mu_{j}(\cdot,\cdot)$ is continuous, we claim that there exists a
 constant $C_{2}>0$ such that
 $$ 
\left|\mu^{+}_{j}(c_0,B,a)-\mu_{j}(c_0,B)\right| \leq C_{2}a
 $$ 
for $0<a<\frac{1}{2\gamma_{+}}$ and $(c_0,B)\in I$. In a similar way, we
 infer the existence of a constant $C_{3}>0$ such that
 $$ 
\left|\mu^{-}_{j}(c_0,B,a)-\mu_{j}(c_0,B)\right| \leq C_{3}a
 $$ 
for $0<a<\frac{1}{2\gamma_{+}}$ and $(c_0,B)\in I$.\\


Let us recall the following result from \cite{EY1}.

\begin{prop}\label{EST2}
(a) Suppose that $\beta a>\frac{8}{3}.$ Then $T^+_{a,\beta}$ has only 
one negative eigenvalue, which we denote by $\zeta_{a,\beta}.$ It 
satisffies the ineqality\\
$\frac{-1}{4} \beta^2 < \zeta_{a,\beta} < \frac{-1}{4} \beta^2 + 
2\beta^2 \exp (\frac{-1}{2} \beta).$  \\
(b) Let $\beta >8$ and $\beta >\frac{8}{3}\gamma_+.$ Then 
$T_{a,\beta}^-$ has a unique negative eigenvalue $\zeta_{a,\beta}^-,$ and moreover, 
we have\\
$\frac{-1}{4} \beta^2 - \frac{2205}{16} \beta^2\exp (\frac{-1}{2} 
\beta)    < \zeta_{a,\beta} ^- < \frac{-1}{4} \beta^2 .$ 

\end{prop}

{\bf{Proof of theorem \ref{t1}}}

 We take
 $a(\beta) =6\beta^{-1}\ln\beta$. Let $\xi^{\pm}_{\beta,j}$ be
 the $j$-th eigenvalue of $T^{\pm}_{a(\beta),\beta}$, by
 Proposition~\ref{EST2} we have
 $$ 
\xi^{\pm}_{\beta,1}=\zeta^{\pm}_{a(\beta)\,,\beta},\quad
\xi_{\beta,2}^{\pm}\geq 0\,.
 $$ 
From the decompositions (\ref{decomp}) we infer that $\{
\xi^{\pm}_{\beta,j} +\mu_{k}^{\pm}
(c_0,B,a(\beta))\}_{j,k\in{\N}}$ properly ordered, is the
 sequence of the eigenvalues of $\hat{H}^{\pm}_{ c_0,B,a(\beta),\beta}$
 counted with multiplicity. Propositions~\ref{EST1} gives
 \begin{equation} \label{lowest}
\xi^{\pm}_{\beta,j}+\mu_{k}( c_0,B,a(\beta))\geq \mu_{1}^{\pm}(
c_0,B,a(\beta))=\mu_{1}(c_0,B)+\mathcal{O}(\beta^{-1}\ln\beta)
 \end{equation}
for $(c_0,B)\in I$, $j\geq 2$, and $k\geq 1$, where the error term is
 uniform with respect to $(c_0,B)\in I$. For a fixed $j\in \N$, we
 take
 $$ 
\tau^{\pm}_{c_0,B,\beta,j}=\zeta^{\pm}_{a(\beta),\beta}
+\mu_{j}^{\pm}(c_0,B,a(\beta)).
 $$ 
Combining Propositions~\ref{EST1} and \ref{EST2} we get
 \begin{equation} \label{tauasympt}
\tau^{\pm}_{c_0,B,\beta,j}=-\frac{1}{4}\beta^{2}
+\mu_{j}(c_0,B)+\mathcal{O}(\beta^{-1}\ln\beta)\quad\mathrm{as}
\quad\beta \to\infty\,,
 \end{equation}
where the error term is uniform with respect to $(c_0,B)\in I$. Let us
 fix $n\in\N$. Combining (\ref{lowest}) with
 (\ref{tauasympt}) we infer that there exists $\beta(n,I)>0$
 such that the inequalities
 $$ 
\tau^{+}_{c_0,B,\beta,n}<0,\quad
\tau^{+}_{c_0,B,\beta,n}<\xi^{+}_{\beta,j}+\mu_{k}^{+}(
c_0,B,a(\beta)),\quad
\tau^{-}_{c_0,B,\beta,n}<\xi^{-}_{\beta,j}+\mu_{k}^{-}( c_0,B,a(\beta))
 $$ 
hold for $(c_0,B)\in I$, $\beta\geq\beta(n,I)$, $j\geq 2$, and $k\geq
1$. Hence the $j$-th eigenvalue of $\hat{H}^{\pm}_{ c_0,B,a(\beta),
\beta}$ counted with multiplicity is $\tau^{\pm}_{c_0,B,\beta,j}$
 for $(c_0,B)\in I$, $j\leq n$, and $\beta\geq \beta(n,I)$. Let $(c_0,B)\in
I$ and $\beta\geq \beta(n,I)$. We denote by  $\kappa^{\pm}_{j}
(c_0,B,\beta)$ the $j$-th eigenvalue of $L^{\pm}_{c_0,B,a,\beta}$.
 Combining our basic estimate and the resultt of \cite{EY2} with
 Lemmas~\ref{1} and \ref{2}, relations
 \ref{ac1} and \ref{ac2}, and the min-max principle, we
 arrive at the inequalities
 \begin{equation} \label{finalest}
\tau^{-}_{c_0,B,\beta,j}\leq\kappa^{-}_{j}(c_0,B,\beta)\quad
\mathrm{and}\quad \kappa^{+}_{j}(c_0,B,\beta)\leq\tau^{+}_{
c_0,B,\beta,j}\quad\mathrm{for}\quad 1\leq j\leq n\,,
 \end{equation}
so we have $\kappa^{+}_{n}(c_0,B,\beta)<0<\inf
\sigma_{\mathrm{ess}}(H_{c_0,B,\beta})$. Hence the min-max principle
 and and the result of \cite{rs} imply that $H_{c_0,B,\beta}$ has at least $n$
 eigenvalues in $(-\infty,\kappa^{+}_{n}(c_0,B,\beta)]$. Given $1\leq
j\leq n$, we denote by $\lambda_{j}(c_0,B,\beta)$ the $j$-th
 eigenvalue of $H_{c_0,B,\beta}$. It satisfies
$$\kappa_{j}^{-}(c_0,B,\beta)\leq\lambda_{j} (c_0,B,\beta)
\leq\kappa_{j}^{+} (c_0,B,\beta)\quad\mathrm{for}\quad 1\leq j\leq
n\,;$$
this together with (\ref{tauasympt}) and (\ref{finalest})
 implies that
 $$ 
\lambda_{j}(c_0,B,\beta)=-\frac{1}{4}\beta^{2}
+\mu_{j}(c_0,B)+\mathcal{O}(\beta^{-1}\ln\beta)
\quad\mathrm{as}\quad\beta \to\infty\quad \mathrm{for} \quad
1\leq j\leq n\,,
 $$ 
where the error term is uniform with respect to $(c_0,B)\in I$. This
 completes the proof.

{\bf{Proof of corollary \ref{c1}}} 
The theorem\ref{t1} with \cite{rs} (theorem XIII.89)  yields the claim. 

\section{Remarks}

Here, if we take $c_0=0$ we recovered the results of \cite{EY1}. If we take $B=0$, we recovered the results of $\cite{hh}$.

{\bf Acknowledgments}
G. H thanks the Abdus Salam ICTP for derive financial assistance.  M.N.H 
acknowledge the Belgian Cooperation CUD-CIUF-UNB for financial support.

\end{document}